# Antarjami: Exploring psychometric evaluation through a computer-based game

Anirban Lahiri[ζ], Utanko Mitra[κ], Sunreeta Sen[ζ], Mreenal Chakraborty[ζ], Max Kleiman-Weiner[η],
Rajlakshmi Guha[κ], Pabitra Mitra[κ], Anupam Basu[κ], Partha Pratim Chakraborty[κ]

Arndit Ltd. [ζ]　　　Indian Institute of Technology, Kharagpur[κ]　　　Massachusetts Institute of Technology[η]

**Abstract**

A number of questionnaire based psychometric testing frameworks are commonly used in the industry and academia for example OCEAN (Five factor) indicator, MBTI (Myer's Brigg Type Indicator) etc. However, questionnaire based psychometric tests have some shortcomings. This work explores whether some of these shortcomings can be overcome through computer-based gaming platforms for evaluating psychometric parameters.

This paper describes a computer based psychometric game framework called Antarjami. It that has been developed for investigating the feasibility of extracting psychometric parameters through computer-based games and underlying improvements in the area of modern artificial intelligence.

## Background

The workings of each of our minds is unique. Even though we share some characteristics with our peers and our family members, each one of us has a unique combination of traits. These are often manifested by how a person behaves when faced with a particular situation. A number of psychometric frameworks exist for modelling the characteristic traits of a person E.g. OCEAN (five factor) indicator (Goldberg, 1992) MBTI (Myer's Brigg Type Indicator) (Myers, 1980)

In past psychologists have typically used questionnaires to assess the mindset of individuals. Although these have served the purpose for many decades, they have some shortcoming as outlined below:

- Questionnaires designed to be sufficiently detailed so as to identify particular characteristics of individuals would be very long and tedious. Thus, it would be time consuming and boring for any person.
- A person answering a questionnaire often tries to use their logical thinking or slow thinking (Kahneman 2012) or their conscious mind (Westen, 1999). However, in many social situations a person might unknowingly use their subconscious/unconscious mind (Westen, 1999) or their fast-thinking (Kahneman, 2012)
- A human being does not feel the same, every day of their life. How they feel while answering the questionnaire, greatly affects the conclusions from the questionnaire. Also, how would their behaviors change when subjected to stress (Quenk, 2000).
- Questionnaires assume that people are honest when answering questions about themselves. However, what often happens is that the person filling in the questionnaire is influenced by trying to guess what is expected of them. This influence is especially true if the questionnaire is given to them by their supervisor or manager or in some particular context.

Now, how does a person assess the mindset of another person when they meet them for the first time? This can often take fairly long for them. Alternatively, a trained psychologist or psychiatrist can interview the person to gauge their mental preferences.

This work explores whether it is possible for a computer to assess the mindset of a person within a short period of time. This paper describes psychoanalytic game framework that attempts to gauge the mindset of the player. Furthermore, game theory has been used widely in behavioral economics, strategic analysis and other applied sciences. This work applies game theory in the area of psychology to better understand the mindset of human beings. The following sections outline some of the mechanics of the game and how they can reveal psychological traits of the players.

## Prior Work

A number of studies have been done in the past to relate game based behavior to cognitive processes (Bechara, 1994). Other studies have focused on deriving computational models and formalisms of game agent (Max Wiener, 2019). Albrecht and Stone (2017) provides a comprehensive survey of modelling approaches to intelligent agents.

In the real world, game plays or tasks are used to extract behavioral traits especially for judging leadership qualities in the armed forces (Steinberg, 1990)

The authors are not aware of any other work which involved the extraction of psychometric parameters from a computer based game.

## OCEAN - Five Factors

The Big Five personality traits, also known as the five-factor model (FFM), is a taxonomy for personality traits (Goldberg, 1992). It is based on common language descriptors. The five factors have been defined as follows:

- Extraversion: (outgoing/energetic vs. solitary/reserved)
- Agreeableness (friendly/compassionate vs. challenging/detached)
- Openness to experience (inventive/curious vs. consistent/cautious).
- Conscientiousness (efficient/organized vs. easy-going/careless).
- Neuroticism (sensitive/nervous vs. secure/confident).

These factors in the past were analyzed by the personality survey data collected from people. This survey data is mainly derived using question-answer sets and has the drawbacks as outlined in the previous section.

This work attempts to determine the 5 traits of human personality through the Antarjami Game Framework instead of any question sets. The next section outlines the game framework. How the 5 factors can be extracted from

## Details of the Antarjami Game

The Antarjami game framework consists of a sequence of game levels. In each level of the game there is 2-dimensional grid within which the players can navigate as shown in Figure 1. Players are shown as different colored circular objects. By default, the current player who is playing the game (also referred to as the subject throughout the remaining article) is shown using a blue circle. Each level has one or more bubble emitters shown in grey Figure1, which emits a stream of bubbles (also shown in grey) similar to soap bubble blowers that children often play with. The objective of the players is to collect the maximum number of bubbles.

Apart from the player who is playing the game on the computer there are multiple other additional players in each level, shown as circular objects of other colors (as Red, Green, Purple and Black in Figure 1). These additional players are driven by various AI (Artificial Intelligence) Engines/Algorithms and are named by their characteristics as Lazy (moves slowly), Greedy (attempts to take the shortest path towards the emitter), Imitator (copies the moves of the participant), Adaptive (calculates a path to square with the highest bubble floe rate) and Irritator (tries to annoy the participant by getting in their way). These characteristics are not communicated directly to the player/participant and it is up to them to find these characteristics out during the gameplay. These peculiar characteristic traits are somewhat similar to those found in people encountered in the real world (like laziness, copying the moves of subject, greedy behavior etc.) They have been included to draw out the characteristic traits and reactions of the subject to real world like scenarios. More detailed analyses regarding the interactions between subject with the other players would be explained in the remaining paragraphs.

Now, given that the objective of the player/subject is to collect the maximum number of bubbles, he or she can either decide to play the levels by themselves or enlist the help of one or more of the additional players. They are provided with a user-interface a given in Figure 3 to choose their teammates.

If the player teams up with N other players then, 25% of the points that it earns will be deducted from it and would be split equally between the other players that he/she teams up with.

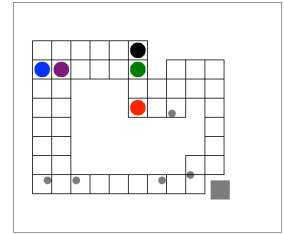

Figure 1: First Level of the Antarjami Game levels.

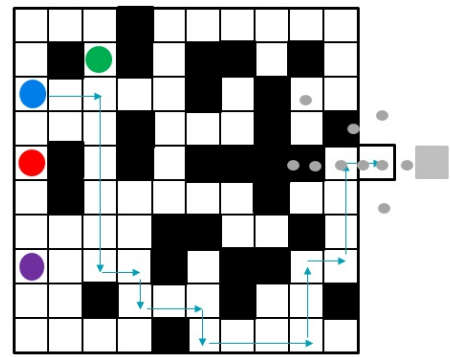

Figure 2: One of the more complex levels.

The other players in the team would also be subjected to the same treatment, and the current player will receive some points from them. Since each of the other players have a particular characteristic throughout the game (based on the AI engine allocated to it), the ability of the subject to select their team is also evaluated in the game.

The game starts with a simple level to familiarize the player with the rules of the game and then progresses to increasingly more complex levels (e.g. Figure 2). In the more complex levels there can be multiple paths and positions which might be pseudo-optimal to reaching a relatively high score.

In addition to the above mechanics of the game, there is also a provision for players to chat to each other, using a chat pop-up in the game (shown in Figure 4). Though this feature in the game is currently quite rudimentary it is a useful tool to judge some of the personality traits.

Though the player is encouraged to achieve a high-score the game actually evaluates both the score and the manner in which the score was achieved to create a map of the player's personality in terms of the five factors.

## Five Factor Evaluation through Game Scenarios

This section explains the 5 factors and how they are evaluated from the game responses. The game creates certain scenarios such that the subject needs to make clearly distinct choices or decisions similar to when answering a questionnaire. However, these decisions are made far more rapidly and in a subconscious fashion without thinking of the actual question regarding their personality traits. Therefore, we hypothesized these decisions to be a truer reflection of the subject's own characteristic traits rather than artificially though out answers. The obtained results and a comparison with scores given by expert psychologists is presented in a later section.

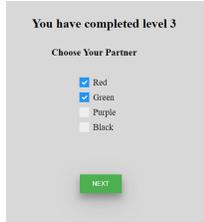

Figure 3: One of the more complex levels.

**Extraversion**

Extraversion refers to the tendency of a person to seek simulation through interaction with others, sociability and talkativeness are signs of this traits. Extraversion is manifested in the game if the player is keen to collaborate with other players and build teams rather than playing as an individual. The subject is judged to me more extrovert particularly if they include a number of non-performing players in their team, this will be explained also in the context of conscientiousness. Players with high extraversion also typically move rapidly during the starting levels of the game. They are also keen to use the chat pop-up window shown in Figure 4.

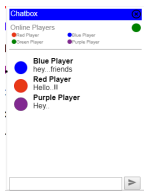

Figure 4: Pop-up Chat window

**Openness to Experience**

This refers to the curiosity of the subject to gain a variety of experiences, openness to adventure, unusual ideas and concepts. In particular levels of the game there are certain greyed out blocks with some hidden bubble emitter (refer Figure 5). These would be discovered by the subject only if they explore these greyed out areas. And in doing so they would also collect extra points in those levels.

In addition, the subjects are asked if they would like to play the game at a more difficult or advanced level. It is found that those who are extremely open to experiences opt to play the game at a more difficult level. A weighted combination of the above metrics indicates how open a subject is to new experiences.

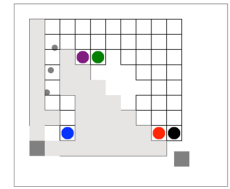

Figure 5: Level showing greyed out areas

**Agreeableness**

Tendency to be friendly, compassionated and cooperative rather than being suspicious or antagonistic towards others is termed agreeableness. In the context of the game, this refers to whether the subject is willing to let other players also score or is actively preventing them from scoring. An example is shown in Figure 6, the positions **a**, **b** and **c** all have similar probabilities for capturing bubbles.

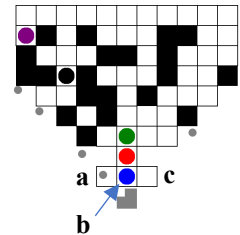

Figure 6: Example of level indicating agreeableness

If the subject occupies position **b** then it prevents the others from scoring while if the subject moves to either position **a** or **c** then the other players can also advance their scores.

**Conscientiousness**

It refers to the personality trait of being careful, diligent, being organized, taking obligations seriously, and the desire to do tasks well. Conscientious people tend to be efficient and organized as opposed to easy-going and disorderly. They exhibit a tendency to show self-discipline, act dutifully, and aim for achievement; they display planned rather than spontaneous behavior.

Within the game framework, conscientiousness is measured by the how well planned and organized the subject is. Certain challenges are presented to them which requires observation and planning before they act. For example, in the level shown in Figure 2, the subject would need to plan out their route before they start moving. The optimal path for collecting maximum number of bubbles in this level is shown by blue arrows. Failing which, the subject has a good chance they would get trapped in one of the other path since one of the other players would block its exit.

How observant the subject is about the other players and how carefully they choose players to join their team is also weighed in. As explained previously not all the AI driven players have the same level of intelligence. In addition, the score sharing mechanism of the game described previously implies that if underperforming players are included in the subject's team then scores would suffer.

A third metric for conscientiousness is as follows. If there are multiple bubble emitters present in a level, then the subject needs to choose a position such that they can benefit from the higher rate bubble emitter.

**Neuroticism**

The tendency to experience unpleasant emotions easily, such as anger, anxiety, depression, and vulnerability. Emotional stability and impulse control are also related to this metric.

The game framework creates some scenario for the subject to potentially get impatient or frustrated. In these scenarios it might still be possible to find alternate solutions if they can put aside their negative feelings. Also, if their negative emotions affect them significantly then their scores for the following level suffer drastically.

An example of such a scenario is shown in Figure 6. The subject is put in a situation where it is unable to collect any bubbles, whereas all the other players reach key positions for collecting bubbles quite quickly. If the subject dwells on negative emotions, then their score for the subsequent level suffers significantly. Alternately, if they have low neuroticism and the continue to look for a solution in this level they would realize that it is possible to move the Purple player away from its position and take its place, since the Purple player is a follower of the subject.

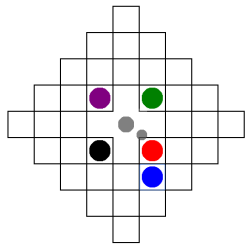

Figure 6: Example of level testing neuroticism

## Methodology

This game-based framework for psychometric evaluation has been evolved and refined across many iterations over a period of more than 2 years. The process is described as follows:
First a simplistic gaming framework was created (inspired by Kleiman-Weiner et. al., 2016) to evaluate the feasibility of psychometric parameters extraction from game analyses. The obtained results looked promising and therefore the team continued further with the study. Each iteration involved the study of 10 to 20 participants, each psychometric parameter was evaluated using each of the 3 methods below:

- A questionnaire/form with 50 questions (Goldberg, 1992)
- Evaluation by trained psychologists
- Evaluation through the Antarjami Game Framework

The psychologists interview each participant and use various techniques especially situational judgement tests (Lievens, 2008) to come up with their scores.

The obtained game results were then compared against those obtained through questionnaires and the scores given by a panel of trained psychologists for individual participants. The differences between each of the 3 methods of evaluation were taken in account and the game framework was updated accordingly. This involved both creating more scenarios as well as updating the methods for computing the psychometric parameters. The final score for each parameter is obtained through a weighted average of individual scenario scores.

In case the data obtained through the questionnaire and that score from the psychologist was in contradiction the score from the psychologist was given priority.

The generic formula for any of the 5 factors as modelled by Antarjami is as follows:

$$f_\eta \left( \frac{1}{S_{Max}} \sum_{i=1}^{\psi} \left[ \frac{S_P}{(1+S_t)^\gamma} \right]^\alpha \left[ 1 - \frac{S_P}{1+S_T} \right]^\beta \tau^\theta \lambda_E \right)$$

$f_\eta$ = Normalization function
$\lambda_E$ = Weight assigned to scenario $i$
$\psi$ = No. of scenarios for the parameter
$S_P$ = Score of Player
$\gamma$ = Influence of team in total score
$\tau$ = No. of team members
$S_T$ = Total score (participant score + team score)
$S_t$ = Team score (obtained by the team selected by participant)
$\alpha, \beta, \theta$ = Exponential Weight Factors

As described below a number of game scenarios are used for analyzing or evaluating any of the 5 factors. The formula shows the relation between these scenarios and the obtained value of the factor. The normalization function gives a relative percentile value with respect to all the participants till date (refer to details in the results section). The weights factors in the formula were tuned using deep learning methods (Deng, 2014) (Durstewitz, 2019). A detailed mathematical treatment of the model is beyond the scope of this paper.

The above process was repeated iteratively with groups of participants, to refine the model (including the weights) and reduce the error margins. It is understandable that there would still be outliers in some cases due to a number of reasons as will be explained in the results and discussion sections.

The Antarjami online game framework can be accessed using the link below:

http://s735719990.websitehome.co.uk/AntarjamiFFI/1.2/beta5/

username: cogsci
password: cogsci2020

The link and the webpage contents have been anonymized to comply with the conference guidelines.

## Results and Discussion

Figure 7 shows the summarized results from a recent batch of 9 participants. These participants were aged between 15 and 23 years. Two of the participants were female while the remaining were male. The sub-plots show a comparison of the results obtained through each of the 3 methods described above. The size of this set is relatively small since the psychologists interviewing the participants need sufficient amount of time for interviewing the candidates so that they can form their opinions and judge the scores. The data corresponding to another larger set of participants is shown later.

It may be observed that in most cases the scores obtained through the game closely reflect those given by the psychologists and are more accurate than the corresponding questionnaire/form based evaluation. The respective correlations between the scores given by Antarjami and the psychologists are as follows: Agreeableness: 0.886, Conscientiousness: 0.735, Openness to Experience: 0.812, Extraversion: 0.595, Neuroticism: 0.654. Even when the score between the psychologists and Antarjami differ, they reflect a similar trend as seen from the plots. In only a few cases do they differ by a sizeable margin since some of these factors are inherently hard to quantify. Also, this happens in relation to conscientiousness and openness to experience, since the game uses a quantitative measure whereas the score from the psychologists is more qualitative in nature.

The results for a second larger dataset is shown in Figure 9. The data corresponds to 47 individuals aged between 15 to 35, with 38 participants in the age group of 15 to 17. Amongst the participant were 17 males and 20 females. A young group of participants were chosen since they are familiar with computer systems and also enthusiastic about playing computer games.

It is apparent from the results that a lot of the participants overestimated their "Openness to Experience". This might be due to the fact some very generic questions are asked in the questionnaire to evaluate this parameter (E.g. "(I) Am full of ideas", "(I) Spend time reflecting on things"). Whereas, in Antarjami it is judged very quantitatively by whether the participant was interested in exploring the additional features of Antarjami. Another major pitfall of the questionnaire was that some pf the participants were not native speakers of English and could not understand some of the questions clearly. Both of these issues of using questionnaires for psychometric evaluation have been addressed by Antarjami.

Antarjami also applies a normalization mechanism to the results obtained through the game by providing a percentile score rather than a pure number. Thereby calibrating all the participants relative to each other. The maximum score ever obtained for a particular parameter on the game is taken as the temporary limit or 100%. Since each human being is unique therefore each of these parameters are evaluated on a continuum rather than being evaluated in a purely binary form (E.g. Introvert vs Extrovert).

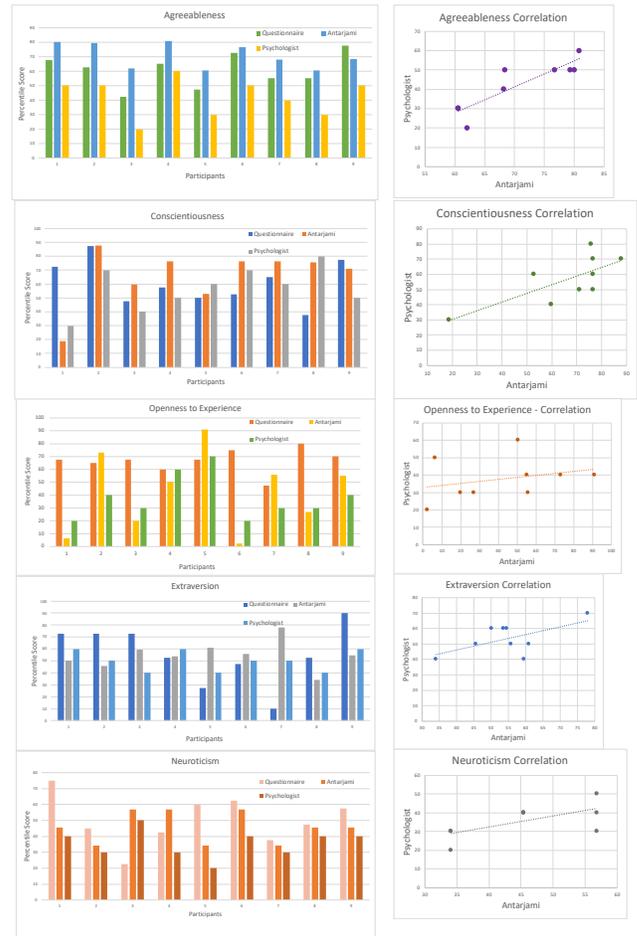

Figure 7: Summarized results

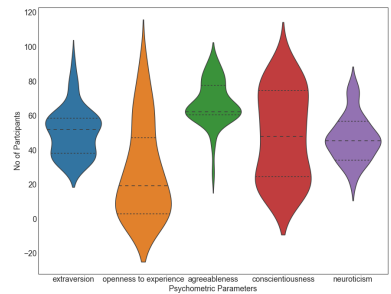

Figure 8: Distribution of the Five factors over the larger sample set of participants

In addition, a sigmoidal mapping function has been used to even out the distributions of each parameter and moderate some of the extreme scores at either end. This, in turn spaces out the intermediate values and improves their information content. Also note that the scores given by the psychologists are multiples of 10 which hints that human evaluators prefer rounded values whereas the percentile scores given by the Antarjami framework can be any value between 0 and 100.

It is also worth mentioning that over the course of the game development across many iterations each participant has only played the game once. This has been done to reduce any artefacts arising from the participant's prior knowledge of the game.

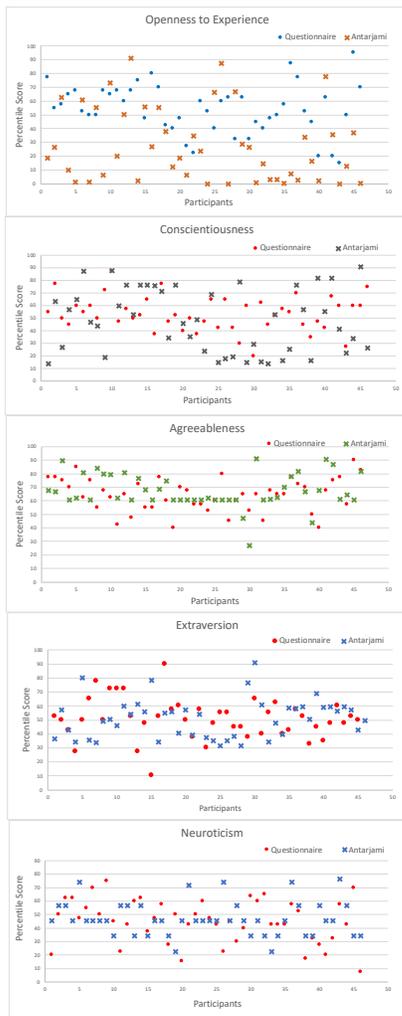

Figure 9: Comparison of the scores obtained through questionnaires and Antarjami.

## Conclusions and Future Work

The Antarjami framework explores the possibility of using computer-based games for psychometric evaluation rather than purely questionnaire based tests. The results obtained so far look promising and display potential of overcoming the shortcomings of questionnaire based test. The framework can also be used to conjunction with questionnaire based tests to gain a more complete personality model of the participants.

There exists a number of avenues for extending Antarjami in the near future. Work is underway to extend Antarjami such that a team of individuals can play on the platform concurrently such that it is possible to observe team dynamics. The Antarjami platform is also being extended with a few other games to form a more complete picture of an individual's personality. Finally, it can also be explored if Antarjami can be extended to evaluate other psychometric parameters like those for MBTI.